\documentstyle[preprint,revtex]{aps}
\begin{document}
\begin{flushright}
\parbox{1.35in}{{\bf CERN-PPE/93-49}\\{\bf MAD/PH/741}\\{\bf FTUV/92-58}
\end{flushright}\par
\begin{title}
{\bf Searching for Exotic Tau Decays}
\end{title}
\begin{center}
\author{R. Alemany}
\begin{instit}
Institut de F\'{\i}sica d'Altes Energies.\\
Universitat Aut\`onoma de Barcelona. 08193 Bellaterra.
\end{instit}
\author{J.J. G\'omez-Cadenas}
\begin{instit}
CERN, CH-1211 Geneva 23
\end{instit}
\author{M.C. Gonzalez-Garcia}
\begin{instit}
Physics Department, University of Wisconsin,\\
Madison, WI53706, USA\\
\end{instit}
and \\
\author{J.W.F. Valle}
\begin{instit}
 Instituto de F\'{\i}sica Corpuscular - IFIC/C.S.I.C.\\
Dept. de F\'{\i}sica Te\`orica, Universitat de Val\`encia\\
46100 Burjassot, Val\`encia, SPAIN
\end{instit}
\end{center}
\thispagestyle{empty}
\begin{abstract}
We  discuss the potential of $\tau$-charm and B factories
for the search of new physics through the study of
rare $\tau$ decays. We consider decays that involve
the violation of lepton flavour conservation. Such
decays bear a close relationship to the physics of
neutrino mass and the properties of the lepton
sector of the electroweak theory.
\end{abstract}

\newpage

\section{Introduction}
\label{in}
\vskip 0.8cm
To present date the hypothesis that the tau lepton and its neutrino are a
standard lepton doublet is confirmed by all existing data. Beyond this the
structure of the weak interaction of the $\tau$ is still poorly determined
experimentally. From the theoretical point of view the charged leptons could
have physical properties substantially different from those predicted by the
standard model. 
lifetimes, rare lepton and/or CP violating decay modes, weak universality
violating interactions, {\sl etc...} Therefore the $\tau$ lepton is an
extremely clean laboratory to look for new physics through the study of its
decay modes. Two complementary approaches can be followed. One can $precisely$
measure the parameters characterizing the well-known leptonic and hadronic
$\tau$ decays and look for possible deviations of their expected ``standard"
values. Alternatively, one can look for $rare$ $\tau$ decays, not expected or
explicitly forbidden in the context of the standard model.

Both precision measurements and rare decay studies will require not only very
good statistics, but also very good detectors providing excellent particle
identification, $\pi^0$ and neutral hadron recognition and hermiticity. These
requirements bring naturally the idea of a {\sl particle factory}
($\tau-$charm factory, B factory, $Z$ factory) as the ideal place to perform
these studies. These facilities can provide very large samples (of the order
of $10^7$ $\sim$ $10^8$  $\tau$ pairs per year) and project state-of-the-art
detectors. The physics prospects, as well as the challenging detector and
machine requirements for these factories have been extensively discussed in
the literature \cite{tcf1,bf1,lep1}.

In this paper we study the sensitivity to possible phenomena involving ``new
physics" that can be achieved by the study of exotic $\tau$ decays in both the
B factory (BF) and the $\tau-$charm factory (tcF). We will study ``typical"
rare $\tau$ decays such as $\tau \rightarrow e(\mu) \gamma$ and $\tau
\rightarrow e(\mu) \pi^0$, three body decays of the type $\tau \rightarrow lll
{}~(l=e,\mu)$ as well as two body decays involving the emission of light weakly
interacting spin zero bosons, for example $\tau \rightarrow e(\mu) J$, where
$J$ denotes the majoron.

Such decays are predicted in a wide class of extensions of the Standard Model
\cite{perga}. In Section two we present some models where the previous decays
can have relative large branching ratios. The experimental aspects of $\tau$
physics at the tcF and BF factories and details of the Monte Carlo simulation
used for this study are given in Section three. In Section four we study the
limits on the different rare $\tau$ decays. Finally in Section five we present
our conclusions.

\section{Theoretical Overview}

Although in the standard model all leptons flavours are exactly conserved
nothing is sacred about these conservation laws when one considers models
beyond the standard model. Therefore looking for rare lepton flavour violating
(LFV) decays is a way to search for so far unknown interactions. There are
several motivations and ways to extend the standard electroweak theory leading
to such rare tau decays. If we restrict ourselves to models with the standard
gauge structure, lepton flavour violation can arise either from a) the mixing
of heavy leptons in the leptonic charged and neutral current or b) due to the
existence of LFV Yukawa couplings to new scalar particles.

In general models of type a) include neutrino masses and, although model
dependent, the expected LFV rates will generally tend to be very much
restricted due to the smallness of neutrino masses that follows from
laboratory, astrophysics and cosmology \cite{perga}. However it has been
recently noted that flavour violation can occur even when neutrinos  are
``protected'' from acquiring a mass by some symmetry such as the conservation
of the  {\sl total} lepton number (this leaves the door open to the possible
violation of {\sl partial} lepton flavours \cite{BER}). This is the case, for
example if neutral heavy leptons (NHL) exist in nature \cite{BER,CERN,MPLA}.
Rare decays processes associated to the existence of NHL include those shown
in table \ref{NHLtable}.

Supersymmetry, both in its conventional realization, as well as in the
alternative scenarios without R parity conservation \cite{RPP,MASI-ROMA} are
examples of models of type b). It offers a possible origin for new signals in
$\tau$ and $\mu$ decays. One example would be the emission of very light
sneutrinos \cite{KANE}, but this possibility is now ruled out in the minimal
supersymmetric standard model. Moreover, sneutrinos are not expected to be
light in this case. However, the emergence of a massless sneutrino is a
generic feature of a class of $SU(2)\otimes U(1)$ supersymmetric models where
the $R_p$ symmetry is broken spontaneously \cite{ARCA}. In these models one
combination of the sneutrinos remains massless because there must be a
physical Nambu-Goldstone boson associated with the spontaneous violation of
lepton number symmetry. This is called the majoron and denoted by $J$. Recent
precision measurements of the $Z$ width at LEP \cite{LEP1} have excluded
\cite{concha} the simplest version of this model \cite{ARCA} but certainly do
not rule out the idea of a massless isosinglet majoron \cite{MASI-ROMA}. As a
consequence majoron emission in many weak decays such as those in table
\ref{Jtable} is experimentally viable and can, in fact, produce sizeable
effects \cite{Nuria}.

\subsection{Rare Tau Decays and NHLS}

Neutral isosinglet heavy leptons arise in many extensions of the electroweak
theory. Being isosinglets, they couple to the standard gauge bosons only
through mixing with ordinary isodoublet neutrinos. The resulting charged
current (CC) leptonic weak interaction  has the form \cite{2227}
\begin{equation}
\L=\frac{g}{\sqrt{2}} W_\mu
\sum_{ij}
\overline l^-_{Li}  \gamma^\mu (K_{Lij}\nu_{Lj}
+K_{Hij}N_{Lj}) + h.c.
\end{equation}
Similarly the neutral current  (NC) Lagrangian for the neutral
leptons is given by
\begin{equation}
\L=\frac{g}{\cos\theta_W}Z_\mu
\sum_{ij}
[\overline \nu_{Li}  \gamma^\mu P_{LLij}\nu_{Lj}+
\overline\nu_{Li} \gamma^\mu P_{LHij} N_{Lj} +
\overline N_{Li} \gamma^\mu P_{HLij} \nu_{Lj} +
\overline N_{Li} \gamma^\mu P_{HHij} N_{Lj}]
\end{equation}
where $P_{AB}=K_A^\dagger K_B$. The admixture in the CC and NC leads in
general to violations of universality which limit the attainable values of
the $K_H$ matrix elements.

These neutral heavy leptons can engender several LFV decays of the charged
leptons, such as those in table \ref{NHLtable}. These are one loop processes
where there is at least one virtual NHL in the loop. For detailed expressions
of the different decay widths see Ref \cite{MPLA}. The widths for the
different processes can be written in the general form
\begin{equation}
\Gamma_p = C_p \times m_\tau^n |\sum_a K_{H\tau a}^* F_p(M_a) K_{H a j} |^2
PS
\end{equation}
where $C_p$  is a constant that depends on the process, $n=3,5$ also depending
on the process. The index $j$ labels the flavour of the final lepton and
$F_p(M_a)$ is the form factor from the integration of the corresponding loop
diagram with a NHL of mass  $M_a$ in some internal leg. Finally PS is the
phase space factor which includes the effect of the masses in the final state.
Therefore the size of the attainable branching ratio will depend on the NHL
mass and of course on the allowed values for the couplings $K_H$. In some
models the $K_H$ are severely constrained by the limits on neutrino masses.
However, it is possible to avoid altogether these constraints. In fact flavour
violation can occur with strictly massless neutrinos. In this case the
constraints on $K_H$ are only those that follow from weak universality,
leading to branching ratios as large as those in Fig. \ref{NHLFig}. We see
that present constraints allow decay branching ratios as large as
$O(10^{-6})$. A detailed discussion of this is given in ref.
\cite{CERN,perga}.

\subsection{Rare Decays and Supersymmetry}

Supersymmetry can produce rare tau decays with detectable rates. This may
occur in simple extensions of the minimal supersymmetric standard model
\cite{SUSYLFV}.

A particularly attractive case correspond to the situation where supersymmetry
is realized in the context of the standard model gauge group in such a way
that R parity is broken spontaneously at the $TeV$ scale. As pointed out these
models contain a massless Nambu-Goldstone, the majoron. In the minimal SUSY
model with broken R parity \cite{ARCA} the existence of the majoron is
accompanied by another light scalar particle, denoted $\rho$, which receives a
mass of order $v_L$, the scale characterizing the spontaneous violation of R
parity. The value of this scale $v_{L}$ is severely constrained by the
astrophysics of stellar cooling and evolution \cite{KIM} to be $v_L < O
\:(30\:KeV)$. As a result there is a new kinematically possible decay mode for
the neutral gauge boson $Z \rightarrow \rho + J $. This leads to an additional
contribution to the invisible $Z$ width. The precision measurements of the $Z$
width at LEP \cite{LEP1} is sufficient to exclude this model \cite{concha}. It
is however possible to formulate the R parity breaking model in such a way
that the majoron is replaced by an $SU(2) \otimes U(1)$ $singlet$ so that it
does not couple to the $Z$ \cite{MASI-ROMA}. The scale characterizing R parity
breaking is now large, {\it i.e.,} $v_{R} = O(1)$ TeV.

In these models the spectrum of the tau decay is modified by single majoron
emission. This can be described in terms of the effective Lagrangian
interaction of the majoron with charged leptons which can given as
\begin{equation}
i J \bar{l}_j \{ \eta_k {\rm L} A_{kj} - \eta_j {\rm R} A_{jk} \} l_k
\label{lag}
\end{equation}
where ${\rm L}$ and ${\rm R}$ are the chiral projectors and $\eta_i$ and
$\eta_j$ are sign factors related to the relative CP parities of the leptons.
The coupling matrix $A$ depends on unknown model parameters and are
constrained by different sets of experimental data as detailed in Ref
\cite{Nuria}.

The corresponding charged lepton decay width is given as
\begin{equation}
\Gamma = \frac {m_k} {32 \pi} [A_{kj}^2 + A_{jk}^2]
\end{equation}
and from this the single majoron emission branching ratio
is easily determined.

As an example we show in Fig. \ref{JFig} the attainable values of the single
majoron emission tau decay branching ratios $\tau\rightarrow \mu J$ as a
function of the tau neutrino mass. This indicate that the flavour-violating
$\tau$ decay processes with single majoron emission could lead to observable
effects for a wide range of the allowed parameters, even at a level similar to
the present experimental sensitivities \cite{SINGLE}.

\section {Tau Physics at the tcF and the BF Factories}

Both tau and B factories may provide complementary information on several
issues on tau physics, including studies of rare tau decay branching ratios.

\subsection{Tau Studies at tcF}

The distinctive feature that  makes the tcF extremely well suited for almost
all $\tau$ physics, is its ability to take data at several operating points
(see Figs. \ref{tauer} and \ref{taucr}) with unique properties:
\begin{itemize}

\item
$3.56 GeV$. Right below $\tau^+\tau^-$ threshold. Taking data at this point
one $measures$ experimentally all non-$\tau$ backgrounds.
\item
$3.57 GeV$.  At this point the $\tau$'s are produced practically at rest. Due
to the Coulomb interaction between the $\tau^+$ and the $\tau^-$, the
$\tau^+\tau^-$ cross-section has a finite value of 0.24 nb, ignoring
initial-state radiation \cite{volo}. With a collision spread of about 1 MeV
the $\tau^+\tau^-$ cross-section is 0.47 nb and $\beta_\tau~=~0.033$. Then,
the two-body $\tau$ decays, such as $\tau\rightarrow\pi\nu_\tau$ and
$\tau\rightarrow K\nu_\tau $ (and also possible exotic decays of the type
$\tau\rightarrow e (\mu) J$, see section two), give rise to monochromatic
secondaries, as illustrated in Fig. \ref{mono}.a. The consequences are clean
signatures for event selection, as well as $kinematic$ separation of the
different decay modes. If a monochromator scheme at the interaction point is
implemented \cite{monocrom} the spread in the collision energy could be
reduced to 0.1 MeV and thus $\beta_\tau~\sim~0.01$, giving rise to still
sharper monochromatic distributions.
\item
$3.67 GeV$.  This energy provides the highest $\tau^+\tau^-$ cross-section
below the  $\psi'(3.69)$ and $D\bar{D}$(3.73) thresholds. Tau decays can be
characterized taking advantage of the fact that they are the only source of
prompt leptons and neutrinos.
\item
$3.69 GeV$.  The highest $\tau^+\tau^-$ cross-section occurs at this energy.
Experiments with a very clean signal (such as the neutrinoless $\tau$ decays
that we will discuss later) can be made here. Notice that, while the standard
optics will result in an observed cross-section of about 4.5 nb, the
monochromator scheme would raise it to 10 nb.
\item
$4.25 GeV$. This is near the maximum energy accessible by tcF. the
$\tau^+\tau^-$ cross section is large and the $\tau$'s boost is moderate,
$\beta_\tau~=~0.54.$ This point is well suited for measurements of the
energy-energy correlations between $\tau$ decay products.
\end{itemize}

The fact that $\tau$ decays are the only source of prompt leptons and
neutrinos at energies below charm threshold makes possible to $single-tag$
$\tau^+\tau^-$ events. Single tagging requires a clean signature from a single
$\tau$ decay. The absence of heavy flavor contamination allows several
distinct signatures: $e+E_{miss},\mu+E_{miss}$ and $\pi+E_{miss}$ (at 3.57
GeV).  Studies \cite{jasp1,jjlfv} indicate backgrounds of between $10^{-3}$
(at 3.57 GeV) and $10^{-4}$ (at 3.67 GeV) in the case of the $e+E_{miss}$
trigger, which has a $\tau^+\tau^-$ event detection efficiency of 24 \%.  %

\subsection{Tau studies at BF}

At the BF, the $\tau$'s are produced with a large boost, $\beta_\tau=0.93$ at
10 GeV. This results in a topological separation of the $\tau$ decay products.
The signature of a $\tau$ pair event is two small-multiplicity (up to five
charged tracks) back-to-back jets. The event is also characterized by missing
energy and transverse momentum due to the undetected neutrinos. $\tau$ physics
is more difficult here than at the tcF. Backgrounds are higher and must be
estimated from Monte Carlo calculation. Single-tagging of $\tau$'s is not
possible. While this is a serious drawback for studies requiring a very
accurate control of systematics (such as the limits on the $\tau$ neutrino
mass or the precision measurement of the $\tau$ topological branching ratios)
there is still a large potential for $\tau$ physics. In particular, the large
boost allows a measurement of the $\tau$ lifetime. Most exotic decays can also
be well studied at the BF.

In fact, one can obtain a good idea of what kind of $\tau$ physics results can
be produced by the BF by looking at the CLEO-II/CESR experiment, with more
than two million $\tau$ pair events recorded.  Recent results from CLEO
\cite{richard} are the measurement of the $\tau$ lifetime (with a relative
error of 2.6 \%, about the same that LEP experiments), the measurement of the
electronic branching ratio (where they achieve a relative error of 3.7 \%
again on the same range than LEP experiments) and the measurement of branching
ratios to several hadronic final states containing one or more $\pi^0$'s.

\subsection{ The Detector for tcF and BF}

Both the tcF and the BF project state-of-the-art detectors \cite{jasp2,bf1}.
Their main features are very similar. Again, the CLEO-II \cite{cldet} detector
is a good example of an operating detector on the line of the requirements of
the tcF and BF. This requirements can be summarized as follows.
\begin{itemize}
\item
Precise charged particle resolution: This is achieved by a precise tracker
(drift chamber) embedded in a high magnetic field (around 1-1.5 T).
\item
Maximum possible solid angle coverage (90-95 \% of the solid angle) for the
electromagnetic calorimetry.
\item
High resolution electromagnetic calorimetry. Crystals like CsI(Tl) and CsI(Na)
are being studied.
\item
Hermetic calorimetry; an outer hadronic calorimeter to detect neutrons and
$K^0_{L}$'s and a careful design to avoid blind regions of the detector.
\item
Good particle identification for $\pi K p$ and $e\mu$ separation; the different
methods being investigated include the use of TOF, dE/dx and RICH.

\end{itemize}

An illustration of these detectors can be found in references
\cite{jasp2,bf1}.

\subsection{Monte Carlo simulation}

Since the general features of the detectors for tcF and BF are similar, it is
an interesting exercise to compare the possibilities offered by both machines
for the study of the $\tau$ exotic decays by assuming the same detector
parameters. In table \ref {pdet}  the basic parameters  of  the  projected tcF
detector are show. We have developed a simplified simulation of this detector
including track momentum smearing in the tracker, electron and photon energy
smearing in the electromagnetic calorimeter, $\gamma-\pi^0$ separation and
detector inefficiencies. Notice that to understand the possibilities offered
by both the tcF and the BF in the search for new physics through the study of
exotic $\tau$ decays a very detailed simulation is not essential, since
straight consequences can be drawn based on very simple facts like the
kinematical separation of $\tau\rightarrow e\nu_e\nu_\tau,\tau\rightarrow
\mu\nu_\mu \nu_\tau$ and $\tau\rightarrow\pi\nu_\tau$ in the tcF or the
excellent mass resolution of both BF and tcF that will allow very easy
identification of $\tau$ neutrinoless decays.

The $\tau^+ \tau^-$ events were generated using the KORALB program.  This
generator includes all effects due to finite mass of the $\tau$, longitudinal
and transverse spin effects of $\tau$'s and e's and order $\alpha^3$ QED
radiative corrections including single hard bremsstrahlung \cite{koralb}. We
have generated and passed through the simulated detector large sample of
events at  two  different center of mass energies;  $\sqrt s~=~3.67$GeV, tcF
and $\sqrt s~=~10$GeV, BF. Table \ref{ntaus} shows the expected number of
$\tau$ pairs per year run at these energy points. A peak luminosity of
$10^{33}$ $cm^{-2} s^{-1}$ is assumed for the tcF. For the BF we assume a
``pessimistic" peak luminosity of  $10^{33}$ $cm^{-2} s^{-1}$ and an
``optimistic" peak luminosity of $10^{34}$ $cm^{-2} s^{-1}.$

\section{Limits on Tau Exotic Decays}
\subsection {$\tau$ selection at the tcF and BF}

Although the selection of $\tau$ events depend to some extent on the physical
channel that one is studying, the general philosophy is clear in both tcF and
BF.
\begin{itemize}
\item
$tcF.$   A $\tau$ event is $single~tagged$ with lepton+$E_{miss}$ (50 $\%$
efficiency) and $\pi+ E_{miss}$ ( at 3.57 GeV with 10 $\%$ efficiency). These
tags provide $\tau$ samples virtually background-free below the charm
threshold and with very low background  above.
 \item
$BF.$  A $\tau$ event is characterized by requiring a two back-to-back jets
with $1.vs.1$ or $1.vs.3$ charged tracks topology.  Events where both tracks
are identified as electrons or muons are rejected to prevent Bhabha and dimuon
backgrounds. In addition, missing energy and/or missing transverse momentum
can be used as a distinctive feature of tau decays on the one-prong side.
Typically one would require the momentum of the one prong side to be smaller
than a fraction of the center of mass energy (i.e., $0.5E_{cms}$).  The
efficiency of these selection criteria is high, around $90\%$, and the
background moderately small,  $3-5\%$.
\end{itemize}

\subsection {$\tau \rightarrow e(\mu) J$}

The two-body decay $\tau\rightarrow e (\mu) J,$ where  $J$  is a massless,
weakly interacting, spin zero Goldstone boson is a very clear example of the
advantage  of being able to produce the $\tau'$s at rest. When $\beta_\tau
\rightarrow 0,$ this decay has the striking signature of a monochromatic
lepton right at the end-point of the standard $\tau\rightarrow l \bar{\nu_l}
\nu_\tau$ momentum distribution. At higher energies, the sensitivity is much
weaker, since the lepton spectrum from the $e(\mu) J$ decay is broad,
spreading over the full spectrum of the standard $\tau\rightarrow l \bar{\nu_l}
\nu_\tau$ decay. See Fig. \ref{mono} (the $\pi$ spectrum from the decay
$\tau\rightarrow\pi\nu_\tau$ is identical to the lepton spectrum from  the
decay $\tau\rightarrow e (\mu) J$).

The procedure to extract a limit on the branching ratio of a possible exotic
decay $\tau\rightarrow e (\mu) J$ at the tcF uses the fact that one knows the
exact shape of both the exotic decay $\tau\rightarrow e (\mu) J$ being
searched and the standard leptonic decay $\tau\rightarrow l \bar{\nu_l}
\nu_\tau$  distributions; this is illustrated in Fig. \ref{exotcf} where
simulated data in which the standard leptonic decay $\tau\rightarrow l
\bar{\nu_l} \nu_\tau$ (18 $\%$ branching ratio)  is mixed  with 1 $\%$
branching ratio exotic decay $\tau\rightarrow e (\mu) J$ is shown. The center
of mass energy is $3.57$ GeV (standard optics is assumed). The region of
interest is $E/E_{max}~>~0.95$. In this region, the shape of the new decay is
known from the data by fitting the $\tau\rightarrow\pi\nu_\tau$ distribution
(in the case of $\tau \rightarrow e J$ one needs to correct for the stronger
effect of the radiative corrections) and also the shape of the standard signal
is known (being simply a straight line). Thus, one can perform a maximum
likelihood fit. This is customarily done by minimizing the quantity
\begin{equation}
 -log( L(\alpha))=-\sum_{i=1}^n log[(1-\alpha)U_s(x_i)~+~\alpha U_e(x_i)],
\end{equation}
where the $n$ is the total number of events passing the energy cut,
$x~=~E/E_{max}$ and the parameter $\alpha$ gives the mixing between the
standard and the exotic decay, whose shapes are described by the functions
$U_s$ and $U_e$ respectively.

At $\sqrt s=~3.57$ GeV with standard optics we obtain a branching ratio
sensitivity to the exotic decay $\tau\rightarrow e (\mu) J$ of $10^{-5}$,
assuming that one can separate the lepton produced in the decay
$\tau\rightarrow e (\mu) J$ and the $\pi$ produced in the decay
$\tau\rightarrow\pi\nu_\tau$. The required $\pi$/l rejection is $10^{-4}.$
This can be  achieved in the case of electrons, where a $e/\pi$ rejection  of
$0.1\%$  (for 0.9 GeV particles) should come from calorimetry and, a further
rejection smaller of 10 $\%$ from TOF and dE/dx. A RICH \cite{ribe} would
further improve the $e/\pi$ rejection by at least a factor $10^{-2}.$ On the
other hand, $\mu/\pi$ separation is limited to the order of 1 $\%$ unless a
RICH  is used. Thus, one could obtain the limits, $B(\tau \rightarrow e
J)~<~10^{-5}$ (standard optics), $B(\tau \rightarrow e J)~<~10^{-6}$
(monochromator); and $B(\tau \rightarrow \mu J)~<~10^{-3}$  (limited by
$\mu/\pi$ separation, it can be improved by an order of magnitude with a
RICH).

The situation is far more difficult at the BF. Here, the leptons arising from
the standard and the exotic decay are mixed up in the full spectrum.
Compare Figs. \ref{exotcf} and \ref{exobf}. The mixing of
exotic/standard decays is the same as above, but now the center of mass energy
is $10.0$ GeV. The best limit that we obtain in this case is
$B(\tau\rightarrow e (\mu) J)~=~5 \cdot 10^{-3}$.

\subsection{Neutrinoless Decays}

These kind of rare $\tau$ decays are characterized by $zero$ missing energy,
and the possibility of fully reconstructing the $\tau$ mass. As a consequence
the event signature is straightforward. One $\tau$ is required to decay to a
clean ``tag" ( i.e. $\tau\rightarrow l \bar{\nu_l} \nu_\tau$ giving a prompt
lepton and missing transverse momentum) and the other to decay to the rare
decay under consideration. This means in general the identification of all or
same of the final state particles, and the reconstruction of the $\tau$ mass.
Since the mass resolution of the tcF or BF detector is very good, the signal
is very clear and the backgrounds small,  in both tcF and BF. As an example we
discuss here two ``typical" rare decays.  A two-body, neutrinoless decay $\tau
\rightarrow \mu \gamma$ and a three body neutrinoless decay $\tau \rightarrow
\mu\mu\mu.$ Clearly many rare decays can searched at both the tcF and the BF
but the general lines will be similar to the ones we discuss here. On the
other hand, the fact that both tcF and BF can reach $10^8$ $\tau$ pairs, and
thus, limits $a~priori$ of $10^{-8}$ for the rare decays that we are
considering, makes it interesting to consider whether backgrounds can be
suppressed to this very low level. We will consider two energies for our
study; $\sqrt s=3.67 $ GeV (tcF  below charm threshold ), and $\sqrt s=10.0$
GeV (BF). The main difference between this two cases is the possible impact of
the backgrounds due to heavy quarks. At both points, with ``standard"
assumptions one gets at least $10^7$ $\sim$ $10^8$ $\tau$ pairs a year, so a
total yield of $10^8$ $\tau$ pairs can be assumed.

To gain a feeling of the rejection to different backgrounds  we compute the
rejection factor for each cut independently. Then, we make the naive approach
that the total rejection factor is the product of the rejection factors for
each cut, ignoring correlations between them. This is necessary in order to
keep the Monte Carlo generation of background events down to some sensible
level and we have checked that the  results are consistent with an exact
calculation using $10^5$ hadronic events.

\subsubsection {Tagging of tau neutrinoless events}

Tau decays have a characteristic signature that allows to tell them apart from
backgrounds very efficiently at both BF and tcF. For the examples that we are
considering (neutrinoless decays with at least a $\mu$ on the final state) we
will select the events by imposing one $\tau$ to decay via the complementary
lepton tag, i.e. the decay $\tau\rightarrow e\nu_e\nu_\tau$. For the
neutrinoless decays with electrons on final state one uses $\tau\rightarrow
\mu\nu_\mu \nu_\tau$ as a tag. Of course, others tags are possible (it also
depends on the specific final state that one is considering) and can result in
increased efficiency but its sensitivity to backgrounds could also be higher
(i.e., tagging the decay $\tau \rightarrow \mu \gamma$ with the decay
$\tau\rightarrow e\nu_e\nu_\tau$ biases against radiative electron and muon
pairs). Thus, for the shake of clarity we take the `gold plated' case and
study whether one can reject backgrounds to the very low level required in
this case.

The event is then accepted if:
\begin{itemize}
\item An energetic electron
(more than 0.4GeV at  tcF and more than 0.8GeV
      at  BF) is identified in the event.
\item The event has missing energy (more than 0.8GeV at tcF, more
      than 1.0GeV at BF).
\end{itemize}

Figs. \ref{tage} and \ref{tagmis} illustrate the effectiveness of the tags. In
Fig. \ref{tage} the energy spectrum of the identified electron is shown for
both the signal (a tau neutrinoless decay with the other tau decaying to an
electron) and the most important background, the multihadronic events. Notice
the logarithmic scale. A cut $p_c$ in the electron momentum of 0.4 $GeV$ at
the tcF (1 GeV at BF) keeps about $70\%$ of the signal while suppressing the
background by more than one order of magnitude. In Fig. \ref{tagmis} the
missing energy spectrum  {\it when an electron has been identified with
$p>p_c$} is shown for signal and background. Cutting at $p_{miss}$ $>$ 0.8 GeV
at the tcF ($p_{miss}$ $>$ 1 GeV BF) again suppresses very efficiently the
background while keeping a high efficiency for the signal.

\subsubsection {$\tau \rightarrow \mu \gamma$}

The signal for  the two energies considered will be a tagged event with only
one additional track and only one photon of at least 100 MeV (multiplicity
cut). This track must be identified as a muon (neutrinoless decay particle
identification cut). Additionally, the total energy of the muon and the photon
must be  within the resolution of the beam energy, labeled $E_{beam}$. The
assumed detector energy  resolutions are 20 MeV for the tcF and 50 MeV for the
BF \cite{jasp2,cinabro}.  Finally the invariant mass of the $\mu-\gamma$
system must be within two $\sigma_m$ (where $\sigma_m$ is the mass resolution
of the tcF or BF, typically 20 MeV) of  the $\tau$ mass (neutrinoless mass
cut) as illustrated in Fig. \ref{masstcf} and in Fig. \ref{massbf}.

The more important backgrounds to consider are multihadronic background, and
internal $\tau$ background ($\tau$ events faking the $\tau \rightarrow \mu
\gamma$ decay). We will consider also $two-\gamma$ events (QED events where
two quasi-real photons scatter to produce a reaction of the type  $e^+e^-
\rightarrow e^+e^-ll (\gamma)$ with $l$ being either an electron or a muon)
although the cross section of the relevant processes (i.e, $\gamma \gamma
\rightarrow ee\tau\tau$ is the only relevant background for the $\tau
\rightarrow \mu\gamma$ decay) is small. Other potentially dangerous
backgrounds like radiative electron or muon pair events are very effectively
suppressed by explicitly biasing against them as explained above   and will
not be considered here, but they must be carefully studied if other additional
tags ($\tau\rightarrow \mu\nu_\mu \nu_\tau
,\tau\rightarrow\pi\nu_\tau,\tau\rightarrow \rho\nu_\tau$)  are used.
\footnotemark
\footnotetext{There are preliminary results from the CLEO collaboration that
are
exploring a wider tagging of exclusive decays.}

Our results are show in table \ref{tmu} and \ref{bmu} for the different
energies considered. As one could expect the most serious background is the
internal $\tau$ background that may be the limiting factor in the sensitivity
for this searches in the limit of very high statistics. The limits on the
branching ratios that can be achieved in a year run  will be: $B(\tau
\rightarrow \gamma\mu) \sim 10^{-7}$ for the tcF and $10^{-6}$ for the BF.

\subsubsection {$\tau \rightarrow \mu\mu\mu$}

This study is very similar to the study described above. The signal for  the
two energies considered will be a tagged event with exactly three additional
tracks (multiplicity cut). At least one  track must be identified as a muon
(neutrinoless decay particle identification cut). The energy of the three
tracks must be close to the beam energy. Finally the invariant mass of the 3
track system must be close to the $\tau$ mass (neutrinoless mass cut) as
illustrated in Fig. \ref{masstcf2} and in Fig. \ref{massbf2}.

Our results are show in table \ref{tele} and \ref{bele} for the different
energies considered. The only limiting factor is the size of the data sample.
Therefore the attainable limits for the branching ratio  for a year run are
$B(\tau \rightarrow \mu\mu\mu) \sim$ $10^{-7}$ for the Tau Charm and the B
Factory.

\section{Conclusions}

We have investigated the potential of $\tau$ and B factories in searching for
new physics through the study of rare $\tau$ decays. Some of these decays can
have substantial branching fractions even in situations where the associated
tau neutrino is very or even massless. Therefore the physics of rare tau
decays is $complementary$ to the efforts of directly improving the neutrino
mass limits, as has been widely advertised can be done at a tau factory.

Our results can be summarized in table \ref{conclu}. In the light of our
results a tcF is preferred for certain rare decays, such as those involving
majoron emission. In this case the attainable branching ratio is limited by
the $e(\mu)/\pi$ separation, for which special attention should be given in
the detector design.  For neutrinoless decays both machines will do very well
and the only limiting factor is basically the size of the data sample.

\acknowledgments
This work was supported by CICYT and the Ministerio de Educaci\'on y  Ciencia
(Spain) and the Department of Energy (USA). We thank J. Kirkby for useful
discussions during this work.

\pagebreak

\figure{Attainable Branching ratio for the $\tau$ decays in the
NHL model consistent with lepton universality. The solid line corresponds to
the decay $l\rightarrow l_i l_j^+l_j^-$, the dashed line corresponds to
$\tau\rightarrow\pi^0 l_i$, the dotted line to $\tau\rightarrow\eta l_i$ and
the dash-dotted line to $\tau\rightarrow l_i \gamma$. In all the cases we have
summed over all possible final leptons.\label{NHLFig}}

\figure{Attainable branching ratios for LFV $\tau$ decays with majoron
emission, as a function of the $\nu_\tau$  mass.\label{JFig}}

\figure{The energy range of the tcF. Plotted is the ratio
$R~=~\sigma(e^+e^- \rightarrow hadrons)/\sigma(e^+e^- \rightarrow \mu^+\mu^-),$
where ``hadrons" include both $q\bar{q}$ and $\tau^+\tau^-$ events. The data
are from DELCO at SPEAR.\label{tauer}}
\figure{The cross section $\sigma(e^+e^- \rightarrow \tau^+\tau^-)$:
a) near threshold (the effects of initial-state radiation are not included);
and b) from threshold to 100 GeV.\label{taucr}}
\figure{The momentum spectra from single-charged-particle $\tau$ decays at
centre-of-mass energies of a) 3.57 GeV  and b) 10 GeV.\label{mono}}
\figure{Searching for the decay $\tau\rightarrow e (\mu) J$ at the tcF. The
points with error  bars are simulated data where the standard leptonic decay
$\tau\rightarrow l \bar{\nu_l} \nu_\tau$ (18$\%$ branching ratio)  is mixed
with 1 $\%$  branching ratio exotic decay $\tau\rightarrow e (\mu) J$. The
solid line superimposed is the standard decay. The center of mass energy is
3.57 GeV and we assume standard optics. \label{exotcf}}
\figure{Searching for the decay $\tau\rightarrow e (\mu) J$ at the BF. The
points with error bars are simulated data where the standard leptonic decay
$\tau\rightarrow l \bar{\nu_l} \nu_\tau$ is mixed  with 1 $\%$ exotic decay
$\tau\rightarrow e (\mu) J$. The solid  line superimposed is the standard
decay. Notice the small deviation near the end-point. The center of mass
energy is 10 GeV.\label{exobf}}
\figure{Electron spectra for $\tau^+\tau^-$ events and electron spectra
(dashed) from QCD events at the center  of mass energy  equal to 3.67 and  10
GeV.\label{tage}}
\figure{Missing energy for neutrinoless decays and QCD background (dashed)
at the center  of mass energy  equal to 3.67 and  10  GeV.\label{tagmis}}
\figure{Invariant mass of the $\tau_1 \rightarrow \mu \gamma$  events,
 $\tau^+\tau^-$ background events, QCD background and two-photon process at
  the center of mass energy equal to 3.67  GeV.\label{masstcf}}
\figure{Invariant mass of the
$\tau_1 \rightarrow \mu \gamma$  events,
 $\tau^+\tau^-$ background events, QCD background and two-photon process at
 the center of mass energy equal to  10.0  GeV.\label{massbf}}
\figure{ a) Invariant mass of the $\tau_1 \rightarrow \mu\mu\mu$  $\tau_2
\rightarrow e \nu \nu$ events; b) Invariant mass of the $e^+ e^- \rightarrow
hadrons$ events; c) Invariant mass of the $\tau^+\tau^-$ events. The center of
mass energy is 3.67 GeV.\label{masstcf2}}
\figure{a) Invariant mass of the $\tau_1 \rightarrow \mu\mu\mu$  $\tau_2
\rightarrow e \nu \nu$ events; b) Invariant mass of the $e^+ e^- \rightarrow
hadrons$ events; c) Invariant mass of the $\tau^+\tau^-$ events. The center of
mass energy is 10.0 GeV.\label{massbf2}}

\begin{table}
\begin{center}
\normalsize
\mbox{  }\\
\mbox{  }\\
\mbox{  }\\
\begin{displaymath}
\begin{array}{|c|c|}
\hline
\mbox{channel} & \mbox{strength} \\
\hline \hline
\tau\rightarrow e\gamma, \mu\gamma       & \sim 10^{-6}\\
\tau\rightarrow e\pi^0, \mu\pi^0         & \sim 10^{-6}\\
\tau\rightarrow e\eta, \mu\eta           & \sim 10^{-7} - 10^{-6}\\
\tau\rightarrow 3e, 3\mu, \mu\mu e, \: etc... & \sim 10^{-7}\\
\hline
\end{array}
\end{displaymath}
\end{center}
\caption{Allowed branching ratios for $\tau$ decays consistent
with lepton universality. The underlying models involve isosinglet
NHLS and the usual neutrinos may be strictly massless}
\label{NHLtable}
\end{table}
\begin{table}
\begin{center}
\begin{displaymath}
\begin{array}{|c|c|}
\hline
\mbox{channel} & \mbox{strength} \\
\hline \hline
\tau\rightarrow \mu J  & \sim 10^{-3}\\
\tau\rightarrow e  J   & \sim 10^{-4}\\
\hline
\end{array}
\end{displaymath}
\end{center}
\caption{Allowed branching ratios for $\tau$ decays in SUSY models with
spontaneously broken R parity}
\label{Jtable}
\end{table}

\begin{table}
\begin{center}
\begin{displaymath}
\begin{array}{|c|c|}   \hline
    & \mbox{tcF Detector}     \\ \hline \hline
   \ \mbox{Charged particles:}                &   \\
     \mbox{Momentum resolution: } \sigma_p/p \:\mbox{(GeV/c)} &
    0.3 \% \oplus 0.3\%/\beta \\
p_{min}^\pi  \mbox{(MeV/c) for efficient tracking} & 50 \\
     \Omega \mbox{(barrel)} (x4\pi \mbox{ str.}) & 90 \% \\ \hline
 \mbox{Photons:}                            &     \\
\mbox{Energy resolution: } \sigma_E/E \mbox{(GeV)}  &
2 \%/E^{1\over{4}} \oplus 1\% \\
\mbox{Angular resolution: } \sigma_{\theta,\phi} \mbox{(mr)} & 5/\sqrt E  \\
E^\gamma_{min} \mbox{(MeV) for efficient detection}  & 10 \\ \hline
\mbox{Particle identification:}     & \\
h \rightarrow e \mbox{ rejection}  & 0.1 \% \\
h \rightarrow \mu \mbox{ rejection}  &  1 \%/p \mbox{(GeV)} + 1 \% \\
K^0_{L}/n \mbox{ detection efficiency}  & 95 \% \\ \hline
\end{array}
\end{displaymath}
\end{center}
\caption{Basic parameters of the tcF detector used in our simulation}
\label{pdet}
\end{table}

\begin{table}
\begin{center}
\begin{displaymath}
\begin{array}{|c|c|c|}    \hline
   \sqrt s \mbox{(GeV)} & \tau^+\tau^- \mbox{ pairs} & \mbox{Machine}     \\
\hline \hline
   3.57     &  0.5~\times~10^7   &  \mbox{tcF (standard optics)}\\
   3.57     &  0.2~\times~10^7   &  \mbox{tcF (monochromator)}\\
   3.67     &  2.4~\times~10^7   &  \mbox{tcF} \\
   3.69     &  5.0~\times~10^7   &  \mbox{tcF (standard optics)} \\
   3.69     &  1.2~\times~10^8   &  \mbox{tcF (monochromator)}   \\
   4.25     &  3.5~\times~10^7   &  \mbox{tcF} \\
   10.4     &  0.9~\times~10^7   & \mbox{BF} (L_{peak}=10^{33}~cm^{-2}
s^{-1})\\
   10.4     &  0.9~\times~10^8   & \mbox{BF} (L_{peak}=10^{34}~cm^{-2}s^{-1})\\
 \hline
\end{array}
\end{displaymath}
\end{center}
\caption{$\tau$ yearly samples at tcF and BF}
\label{ntaus}
\end{table}
\begin{table}
\begin{center}
\begin{displaymath}
\begin{array}{|c|c|c|c|c|}    \hline
\mbox{selection criteria} & \mbox{signal (eff)} & \tau \mbox{bkgnd rej.}
& \mbox{QCD bkgnd rej.} &
\gamma\gamma \mbox{ bkgnd rej.} \\  \hline \hline
\mbox{Tag} &1.5 \times~10^{-1} & 25 \times~10^{-2} &     10^{-3} &
      3 \times~10^{-1} \\
\mbox{Topology} & \sim 1   & 5 \times~10^{-2} & 3 \times~10^{-2} &
      3 \times~10^{-2} \\
\mbox{Inv. Mass} & \sim 1   & 3 \times~10^{-4} & 2 \times~10^{-2} &
      2 \times~10^{-4} \\
E_{beam} & \sim 1   & 2 \times~10^{-2} & 2 \times~10^{-1} &
      5 \times~10^{-3} \\
\mu \mbox{ID}  &\sim 1   &  2 \times~10^{-1} & 4 \times~10^{-3} &
      3 \times~10^{-1} \\ \hline
\mbox{Overall}&1.5  \times~10^{-1}& 1.5   \times~10^{-8}&5. \times~10^{-10 } &
    1.7 \times~10^{-9} \\
 \hline
\end{array}
\end{displaymath}
\end{center}
\caption{ $\tau \rightarrow \mu\gamma$. efficiency and
background rejection at  $\sqrt s =3.67 $ GeV }
\label{tmu}
\end{table}

\begin{table}
\begin{center}
\begin{displaymath}
\begin{array}{|c|c|c|c|c|}    \hline
\mbox{selection criteria} & \mbox{signal (eff)} & \tau \mbox{bkgnd rej.} &
\mbox{QCD bkgnd rej.} &
\gamma\gamma \mbox{ bkgnd rej.} \\ \hline \hline
\mbox{Tag} & 10^{-1} & 15 \times~10^{-2} & 3 \times~10^{-2} &
      4 \times~10^{-1} \\
\mbox{Topology} & \sim 1   & 5 \times~10^{-2} & 4 \times~10^{-3} &
      5 \times~10^{-2} \\
\mbox{Inv. Mass} & \sim 1   & 5 \times~10^{-3} & 2 \times~10^{-2} &
               10^{-3} \\
E_{beam} & \sim 1   & 5 \times~10^{-2} & 5 \times~10^{-3} &
               10^{-3} \\
\mu \mbox{ID}  & \sim 1  &  2 \times~10^{-1} & 6 \times~10^{-2} &
      3 \times~10^{-1} \\ \hline
\mbox{Overall}  & 10^{-1}  & 3.8 \times~10^{-7} & 7.2\times10^{-10} &
      6.0\times10^{-9} \\
 \hline
\end{array}
\end{displaymath}
\end{center}
\caption{ $\tau \rightarrow \mu\gamma$. efficiency and
background rejection at $\sqrt s=10 $ GeV.}
\label{bmu}
\end{table}

\begin{table}
\begin{center}
\begin{displaymath}
\begin{array}{|c|c|c|c|c|}    \hline
\mbox{selection criteria} & \mbox{signal (eff)} & \tau \mbox{bkgnd rej.} &
\mbox{QCD bkgnd rej.} &
\gamma\gamma \mbox{ bkgnd rej.} \\  \hline \hline
\mbox{Tag} & 1.5\times10^{-2} & 2 \times~10^{-1} & 9 \times~10^{-4} &
      5 \times~10^{-3} \\
\mbox{Topology} & \sim 1      & 2 \times~10^{-2} & 5 \times~10^{-2} &
      6 \times~10^{-4} \\
\mbox{Inv. Mass} & \sim 1   & 6 \times~10^{-6} & 9 \times~10^{-3} &
      2 \times~10^{-4} \\
E_{beam} & \sim 1   & 2 \times~10^{-3} & 2 \times~10^{-1} &
      6 \times~10^{-3} \\
\mu \mbox{ID}  &\sim 1   &  1 \times~10^{-2} & 2 \times~10^{-4} &
      1 \times~10^{-1} \\ \hline
\mbox{Overall}&1.5 \times~10^{-2}  &  4 \times~10^{-13} & 1.4 \times~10^{-11} &
      3.6 \times~10^{-13} \\
 \hline
\end{array}
\end{displaymath}
\end{center}
\caption{ $\tau \rightarrow \mu\mu\mu$. efficiency and
background rejection at  $\sqrt s =3.67 $ GeV }
\label{tele}
\end{table}

\begin{table}
\begin{center}
\begin{displaymath}
\begin{array}{|c|c|c|c|c|}    \hline
\mbox{selection criteria} & \mbox{signal (eff)} &\tau \mbox{bkgnd rej.} &
\mbox{QCD bkgnd rej.} &
\gamma\gamma \mbox{ bkgnd rej.} \\ \hline \hline
\mbox{Tag} & 1 \times~10^{-1} & 2 \times~10^{-1} & 3 \times~10^{-2} &
      6 \times~10^{-4} \\
\mbox{Topology} &  \sim  1    & 2 \times~10^{-2} & 3 \times~10^{-3} &
      2 \times~10^{-4} \\
\mbox{Inv. Mass} & \sim 1   & 6 \times~10^{-6} & 2 \times~10^{-2} &
      2 \times~10^{-4} \\
E_{beam} & \sim 1   & 1 \times~10^{-2} & 6 \times~10^{-3} &
      7 \times~10^{-3} \\
\mu \mbox{ID}  & \sim 1   &  1 \times~10^{-2} & 6 \times~10^{-2} &
      2 \times~10^{-2} \\ \hline
\mbox{Overall} & 1 \times~10^{-1}   &  2 \times~10^{-12} & 3.6 \times~10^{-10}
&
      4 \times~10^{-16} \\
 \hline
\end{array}
\end{displaymath}
\end{center}
\caption{ $\tau \rightarrow \mu\mu\mu$. efficiency and
background rejection at $\sqrt s=10 $ GeV.}
\label{bele}
\end{table}
\begin{table}
\begin{center}
\begin{displaymath}
\begin{array}{|c|c|c|}
\hline
\mbox{channel} & \mbox{tcF} & \mbox{BF} \\
\hline \hline  & & \\
\tau\rightarrow e J &
\begin{array} {ccc}   & 10^{-5}  & \mbox{(standard-optics)}\\
                      & 10^{-6}  & \mbox{  (monochromator)}
\end{array}
&  5\times 10^{-3}   \\[0.3cm]
\hline  & & \\
\tau\rightarrow \mu  J  &
\begin{array} {ccc}   & 10^{-3}  & \mbox{      }\\
                      & 10^{-4}  & \mbox{(RICH)}
\end{array}
&   5\times 10^{-3}  \\[0.3cm]
\hline & & \\
\begin{array} {c}
\tau\rightarrow e \gamma \\
\tau\rightarrow \mu \gamma
\end{array}
 &  10^{-7} & 10^{-6}\\[0.3cm]
\hline   & & \\
\tau\rightarrow \mu \mu\mu     &  &  \\
\tau\rightarrow \mu e e     & 10^{-7}   &10^{-7}  \\
\tau\rightarrow  e  \mu\mu     &     &    \\
\tau\rightarrow  e  e e     &    &    \\[0.3cm]
\hline
\end{array}
\end{displaymath}
\end{center}
\caption{Attainable limits for the branching ratios for  different $\tau$
decays in a $\tau$-charm Factory and a B Factory for one year run.
J is  a neutrino like particle (majoron).}
\label{conclu}
\end{table}
\end{document}